\documentclass[oldversion,rnote]{aa}  
\usepackage{graphicx}
\usepackage{txfonts}
\begin{document}
   \title{23 GHz VLBI Observations of SN 2008ax}

   \author{I. Mart\'i-Vidal\inst{1,2}
          \and
          J.M. Marcaide \inst{1}
          \and 
          A. Alberdi\inst{3}
          \and
          J.C. Guirado\inst{1}
          \and 
          M.A. P\'erez-Torres\inst{3}
          \and
          E. Ros\inst{2,1}
          \and
          I.I. Shapiro\inst{4}
          \and
          R.J. Beswick\inst{5}
          \and
          T.W.B. Muxlow\inst{5}
          \and
          A. Pedlar\inst{5}
          \and
          M.K. Argo\inst{6}
          \and
          S. Immler\inst{7}
          \and
          N. Panagia\inst{8,9,10}
          \and
          C.J. Stockdale\inst{11}
          \and
          R.A. Sramek\inst{12}
          \and
          S. Van Dyk\inst{13}
          \and
          K.W. Weiler\inst{14}
          }

   \institute{Dpt. Astronomia i Astrof\'isica, Universitat de Val\`encia,
              C/ Dr. Moliner 50, E-46100 Burjassot, Spain\\
              \email{I.Marti-Vidal@uv.es}
         \and
             Max-Planck-Institut f\"ur Radioastronomie,
             Auf dem H\"ugel 69, D-53121 Bonn, Germany
         \and
             Instituto de Astrofísica de Andalucía (CSIC),
             C/ Camino bajo de Hu\'etor 50, E-18008 Granada, Spain
         \and
             Harvard-Smithsonian Center for Astrophysics,
             60 Garden St., MS 51, Cambridge, MA 02138, USA
         \and
             Jodrell Bank Observatory,
             Macclesfield, Cheshire SK11 9DL, UK
         \and
             Department of Imaging and Applied Physics, 
             Curtin University of Technology,
             Bentley, WA 6845, Australia
         \and
             NASA Goddard Space Flight Center, 
             Astrophysics Science Division, Code 662, Greenbelt, MD 20771, USA
         \and
             Space Telescope Science Institute, 3700 San Martin Drive,
             Baltimore, Maryland 21218, USA
         \and
             INAF - Osservatorio Astrofisico di Catania,
             Via S. Sofia 78, I-95123, Catania, Italy
         \and
             Supernova Ltd, OYV \#131, Northsouth Rd., 
             Virgin Gorda, British Virgin Islands
         \and
             Department of Physics, Marquette University, 
             PO Box 1881, Milwaukee, WI 53201-1881, USA
         \and
             National Radio Astronomy Observatory, 
             PO Box O, Socorro, NM 87801, USA
         \and
             University of California, Astronomy Department,
	     Berkeley, California 94720, USA
         \and
             Naval Research Laboratory,
             Code 7210, Washington, DC 20375-5320, USA
             }

   \date{Accepted in A\&A on 24/03/2009}

 
  \abstract
{We report on phase-referenced 23 GHz Very-Long-Baseline-Interferometry (VLBI) observations 
of the type IIb supernova SN 2008ax, made with the Very Long Baseline Array (VLBA) on 2 April 
2008 (33 days after explosion). These observations resulted in a 
marginal detection of the supernova. The total flux density recovered from our VLBI image is 
0.8$\pm$0.3 mJy (one standard deviation). As it appears, the structure may be interpreted as 
either a core-jet or a double source. However, the supernova structure could be somewhat confused 
with a possible close by noise peak. In such a case, the recovered flux density would decrease to 
0.48$\pm$0.12 mJy, compatible with the flux densities measured with the VLA at 
epochs close in time to our VLBI observations. The lowest 
average expansion velocities derived from our observations are $(1.90 \pm 0.30) \times 10^5$ km s$^{-1}$ 
(case of a double source) and $(5.2 \pm 1.3) \times 10^4$ km s$^{-1}$ (taking the weaker 
source component as a spurious, close by, noise peak, which is the more likely interpretation). 
These velocities are 7.3 and 2 times higher, respectively, than the maximum ejecta velocity 
inferred from optical-line observations.}

   \maketitle


\section{Introduction}

Supernova \object{SN\,2008ax} was discovered in galaxy \object{NGC\,4490} on 3 
March 2008 (Mostardi et al. \cite{Mostardi2008}) at the position 
$\alpha = 12^h 30^m 40.799^s$ and $\delta = 41^{\circ} 38' 14.825''$. 
The host galaxy is $\sim$8\,Mpc distant (de Vaucouleurs 
\cite{Vaucouleurs1976}; we assume an uncertainty of 1\,Mpc for the distance), 
and is dynamically interacting with \object{NGC\,4485}
(forming the pair \object{Arp\,269}). Likely as a consequence of this 
interaction, \object{NGC\,4490} has a relatively high star formation rate 
(Viallefond et al. \cite{Viallefond1980}), which should
results in a correspondingly high supernova rate. 

We assume the discovery date of the supernova was the same as the 
explosion date, since 6 hours before its discovery, the location of the supernova 
was imaged with a limiting magnitude of 18.3 and no emission was 
detected (Nakano \cite{Nakano2008}). Radio emission from SN\,2008ax was monitored 
with the VLA beginning on 3 March 2008, shortly after its discovery (see Stockdale 
et al. \cite{Stockdale2008a}, \cite{Stockdale2008b}), with positive detections made at 
4.9, 8.4, and 22.5\,GHz, beginning on 7 March 2008. These detections were all in the 
millijansky range. 
The early part of the radio light curve of this supernova is qualitatively similar 
to that of SN\,1993J. Since supernova SN\,2008ax was also cataloged as type IIb 
(Chornock \cite{Chornock2008}), as was SN\,1993J, 
there was evidence to believe that the radio emission of SN\,2008ax 
would continue its evolution in a similar way to \object{SN\,1993J}'s. In such a case, 
the flux density of SN\,2008ax should have risen well above the VLBI 
detectability limit near the end of March 2008.

On 11 March 2008, we proposed a {\bf target-of-opportunity} set of global 
VLBI observations of SN\,2008ax at 23\,GHz, in order to detect the supernova 
radio structure and, possibly, its expansion.
Only antennas of the VLBA were allocated and only on 2 April 2008. 
Unfortunately, the flux density {\bf started to drop faster than expected} by that time, 
resulting in only a marginal detection of the supernova. In the next section we 
describe the details of our VLBI observations, and in Section \ref{III} we 
present our results and conclusions.

\section{Observations and Data Reduction}

We observed supernova 2008ax on 2 April 2008, with the 
VLBA (10 identical 25\,m diameter antennas spread over the USA from the 
Virgin Islands to Hawaii). The recording rate was set to 256\,Mbps,
with 2-bit sampling and single polarization mode (LCP), covering a total 
bandwidth of 64\,MHz (8 baseband channels, of 8\,MHz width each). Our 
observations were cross-correlated at the Array Operations 
Center of the National Radio Astronomy Observatory (NRAO) in Socorro 
(New Mexico, USA), using an averaging time of 2 seconds.

The observations of SN\,2008ax were made in phase-reference mode. Each scan of 
the supernova was of $\sim$2 minutes duration, and short observations ($\sim$40 seconds) 
of strong, close by sources were interleaved between these scans of the supernova. 
Since the 23\,GHz flux densities of these close by radio sources 
were unknown at the time of the observations, we chose the two closest, which were also
the strongest at 15\,GHz. Each pass of the 
recording tapes (22 minutes long) was then assigned to one of these two calibrators in 
an alternating scheme. The 12-hour long set of observations could, thus, be 
divided into two sets of roughly equal size. In the first one, we observed the supernova 
using the source \object{J1224+4335} as the phase calibrator (located 2.23 degrees from the 
supernova) and in the second one we used the source \object{J1225+3914} as the phase 
calibrator (located 2.57 degrees from the supernova). 

After the cross correlation, the data were imported into the NRAO Astronomical Image
Processing System ({\sc aips}) for calibration. We performed the amplitude calibration 
using gain curves and system temperatures measured at all antennas. We then used in the
scans of the supernova the time-interpolated antenna gains obtained from hybrid mapping
of the calibrators. 
The phase calibration (with account taken of the structures of the calibrators) was 
performed with standard phase-reference calibration techniques. The data were then 
exported for further reduction in {\sc difmap} (Shepherd et al. \cite{Shepherd1995}). 

\begin{figure}
\centering
\includegraphics[width=9cm]{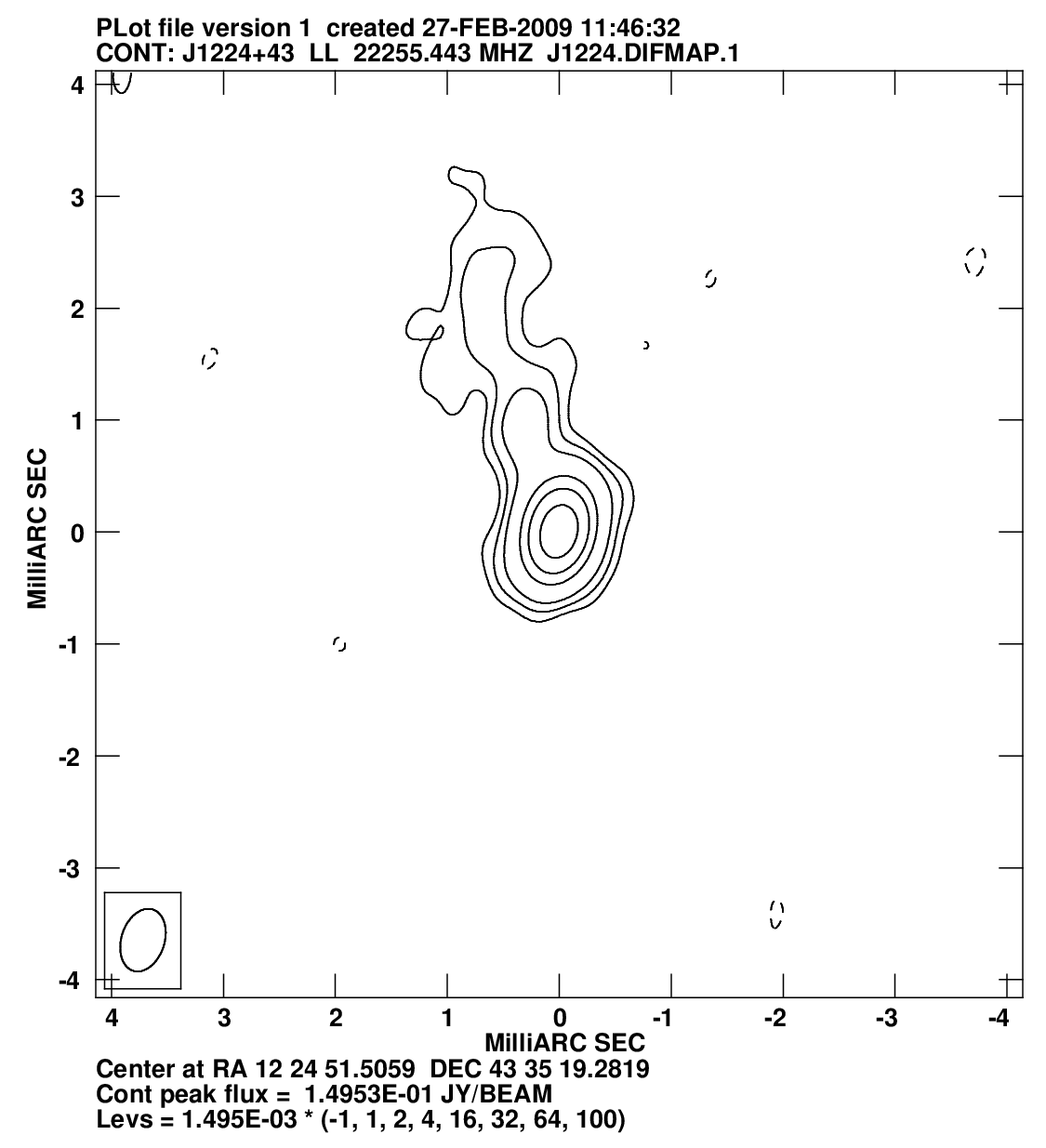}
\caption{Hybrid image of source J1224+4335 obtained from our observations. 
The FWHM of the convolving CLEAN beam is shown at the bottom-left corner.}
\label{Calibrator}
\end{figure}

\section{Results and Conclusions}
\label{III}

The flux density of the calibrator source J1224+4335, obtained from hybrid mapping, 
is 198$\pm$1\,mJy and the flux density of the other calibrator source, J1225+3914, 
obtained with the same procedure, is 122$\pm$1\,mJy. Uncertainties are 3 times the 
root-mean-square, rms, of the hybrid-map residuals (see Readhead \& Wilkinson \cite{Readhead1978}). 
There is no clear detection of SN\,2008ax in any of the phase-referenced images obtained 
using both calibrators. There is not even a clear flux density peak in the 
images obtained from the totality of supernova visibilities (i.e., dynamic range above $\sim$6), 
regardless of the sky coverage of the image or the 
weighting scheme applied in Fourier space. However, when the phase-reference
calibrator J1224+4335 (i.e., the strongest and closest calibrator) is used,
there is a possible detection of the supernova. The detection arises from applying
a visibility weighting in Fourier space with the weight of a pixel proportional to the 
square of the signal-to-noise ratio (SNR) of the visibilities inside that pixel (i.e., 
we increase the array sensitivity in the Fourier inversion). We additionally taper the 
visibilities using a Gaussian, centered at the origin of the uv-plane, with a Full Width at 
Half Maximum (FWHM) of 500\,M$\lambda$. The image obtained with such a visibility weighting 
has a peak flux density located at $0.08$\,mas to the West and $9.2$\,mas to the South of 
the position used at the correlator, which was taken from X-band VLA observations made 
on 8 March 2009. Therefore, the peak flux density detected is located at 
$\alpha = 12^h 30^m 40.79899^s$ and $\delta = 41^{\circ} 38' 14.81580''$, with an 
uncertainty of 0.05\,mas, which is the size of the interferometric beam divided by 
2 times the dynamic range of the image (Thomson et al. \cite{Thomson1986}).
Since this position is based on a phase-reference to J1224+4335 at 23\,GHz, we notice 
that opacity effects in the jet of this source (see Figure \ref{Calibrator}) could 
introduce a systematic shift of several 
mas in the supernova position (the correlation position of J1224+4335 was taken from the 
VLBA Calibrator Survey at 8.4\,GHz; see Beasley et al. \cite{Beasley2002}).
Decreasing the FWHM of the Gaussian taper, or weighting each pixel with a higher power of 
the visibility SNR, results in a slightly better detection of the supernova but, due to 
the large decrease in resolution, at the price of a detailed detection of radio structure.

\begin{figure}
\centering
\includegraphics[width=9cm]{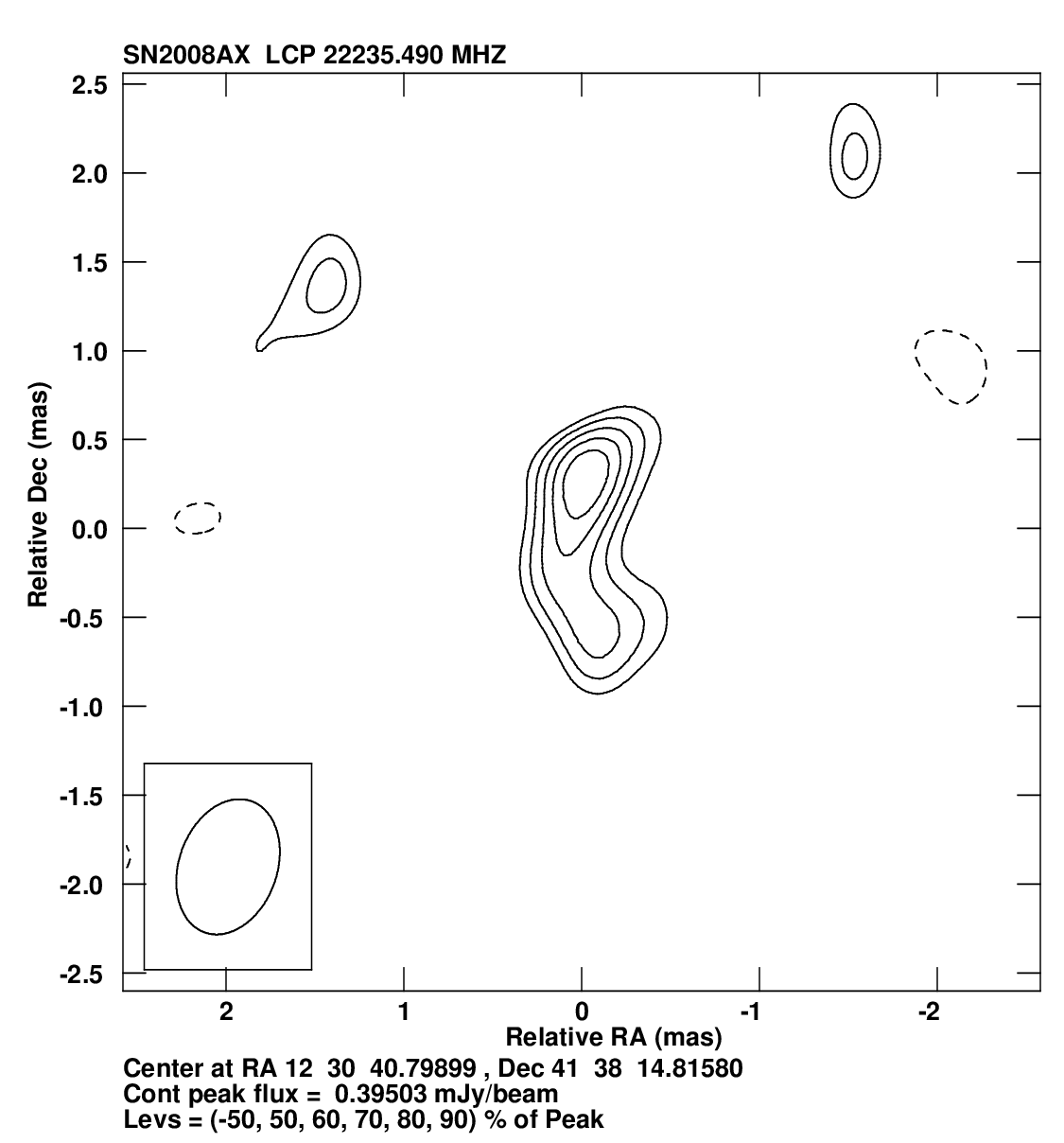}
\caption{CLEAN phase-referenced image of SN\,2008ax (see text). The FWHM of 
the convolving beam is shown at the bottom-left corner.}
\label{SN08ax}
\end{figure}

\begin{figure}
\centering
\includegraphics[width=9cm]{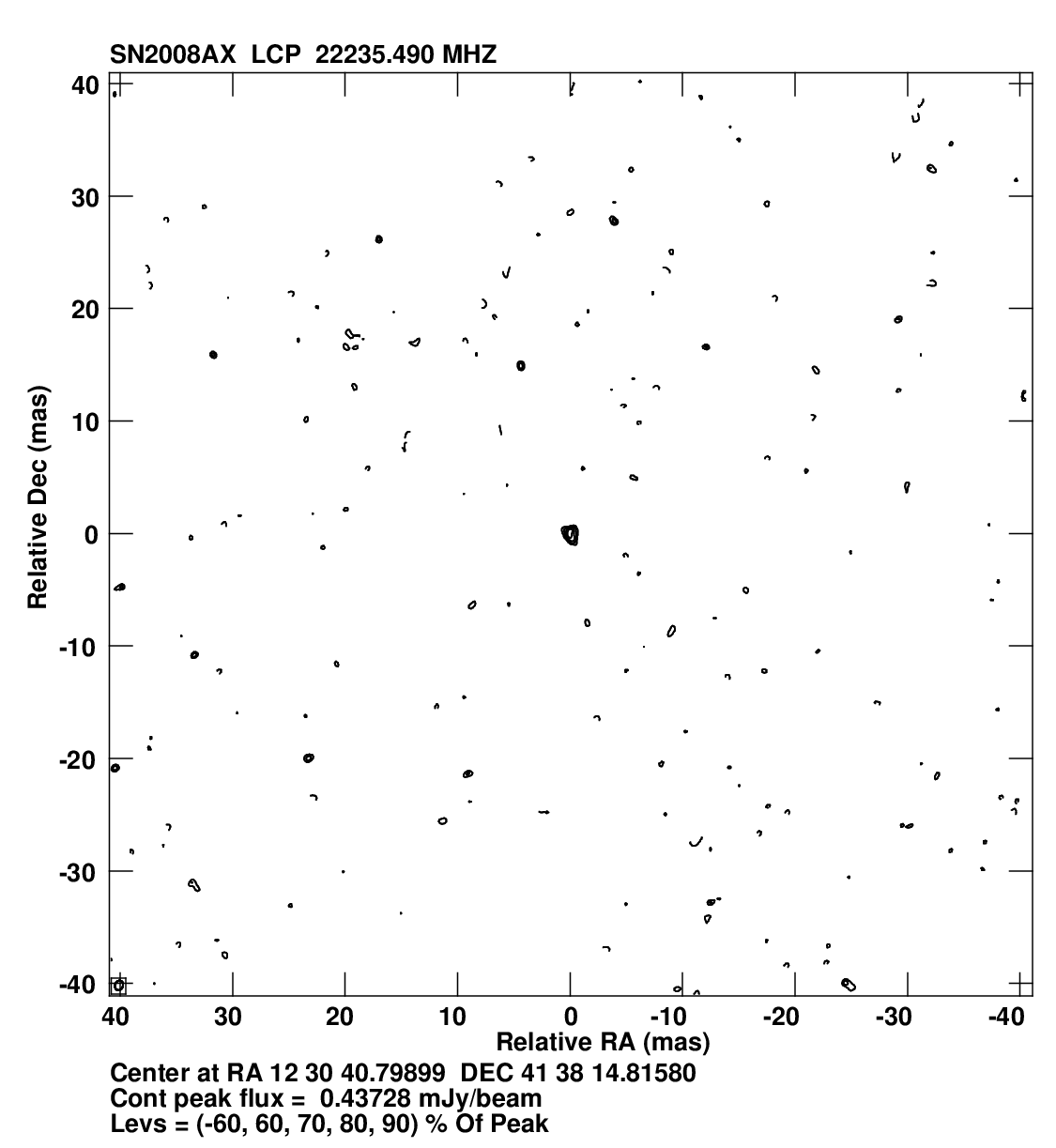}
\caption{Dirty image of SN\,2008ax with a sky coverage of 40$\times$40\,mas. Due to the 
limitations of the FFT algorithm, around 10\% of the visibilities (which correspond to 
the longest baselines) had to be removed in the Fourier inversion for obtaining this 
wide-field image.}
\label{SN08ax-wide}
\end{figure}

After performing a CLEAN deconvolution in the region of the flux density peak, we 
obtain the image shown in Figure \ref{SN08ax}. The rms of the residuals is 
0.087\,mJy\,beam$^{-1}$ and the peak flux density is 0.395\,mJy\,beam$^{-1}$, so the 
dynamic range of the image is 4.7. The total flux density obtained after a 
deconvolution using the CLEAN algorithm is 0.8\,mJy. Remarkably, 
the flux density recovered from our VLBI image is a factor $\sim$1.7 larger than the flux 
densities registered with the VLA by Stockdale et al. (\cite{Stockdale2008b}) at epochs close in 
time to the epoch of our observations ($0.48\pm0.12$\,mJy on April 1, and $0.45\pm0.10$\,mJy on April 3).
This large discrepancy indicates that the source structure shown 
in Figure \ref{SN08ax} may be a chance superposition of a marginal detection (North) and 
a prominent noise peak (South), as we explain below.  
The integrated flux density of each of the noise peaks of the residual image in a 
40$\times$40\,mas square around the source is less than 0.4\,mJy (i.e., less than 50\% 
of the integrated flux density of the source). We show this wide-field image in 
Figure \ref{SN08ax-wide}.

If the structure shown in Figure \ref{SN08ax} is real, does it correspond to 
a core-jet or is it part of a shell? Or, as we suggest above, can this structure be a 
combination of a marginal detection (North) with a stronger noise peak (South) than 
elsewhere in the map? Difficult to say.

\subsubsection*{Case 1. Partial Shell-like Structure}

For the radio structure part of a shell, we can compare its 50\% contour 
level with that of a shell model convolved with the same beam. For a shell model with 
a fractional shell width of 0.3, which is the shell width found for SN\,1993J by 
Marcaide et al. (\cite{Marcaide2009}), the outer radius of SN\,2008ax 
would be 1.15$\pm$0.15\,mas (hereafter, all the uncertainties given are equal to the 
square root of the corresponding diagonal element of the covariance matrix, with the 
errors having been first uniformly scaled so that the reduced
$\chi^2$ is equal to 1). This size translates into an average expansion velocity of 
$(4.8 \pm 0.8) \times 10^5$\,km\,s$^{-1}$, which is superluminal. Indeed, we still obtain 
superluminal expansion velocities if we change the fractional shell width to different, 
unrealistic, values such as 0.1 (a narrower shell width translates into a smaller fitted shell 
size). Hence, it is unlikely that the radio structure is part of an expanding shell. 

\subsubsection*{Case 2. Double Source} 

If we instead fit the visibilities to two point sources, one to model the brightness peak 
(at the North) and the other one to model the source extension towards the South, we find 
components of 0.40$\pm$0.11 and 0.24$\pm$0.11\,mJy, separated by 0.93$\pm$0.10\,mas. This 
result translates into an average 
relative velocity between components of $(3.90 \pm 0.62) \times 10^5$\,km\,s$^{-1}$, which is 
superluminal. If 
the two components were moving in opposite directions with respect to the explosion 
center, the average expansion velocity of the radiostructure would be 
$(1.90 \pm 0.31) \times 10^5$\,km\,s$^{-1}$, a factor $\sim$7.3 higher than the maximum ejecta 
velocity estimated from the optical-line emission of this supernova 
($\sim 2.6 \times 10^4$\,km\,s$^{-1}$, Blondin \cite{Blondin2008}). This velocity is also much 
higher than the typical expansion velocities of the radiostructures of other supernovae 
($\sim 1-2 \times 10^4$\,km\,s$^{-1}$). Hence, the two-point source model is also unlikely.

\subsubsection*{Case 3. Detection with a Close by Noise Peak}

Perhaps, then, we could consider that the structure shown in the map is due to a chance (near) 
superposition of a marginally detected radio source and a noise peak. In this case, the 
radio emission would not be resolved and its detection would be even more marginal. Fitting 
a shell model (with a fractional width of 0.3) to the northern flux density peak results in a 
source outer diameter of $0.25\pm0.05$\,mas and 
a flux density of $0.48\pm0.12$\,mJy. We notice that this flux density is consistent with 
the flux densities registered by Stockdale et al. \cite{Stockdale2008b} at the same radio 
frequency and at epochs enclosing that of our VLBI 
observations. The resulting average expansion velocity is $(5.2 \pm 1.3) \times 10^4$\,km\,s$^{-1}$, 
a factor of 2 larger than the velocity inferred from optical-line emission. This velocity is 
also $\sim$3 times larger than the expansion velocities of the other radio 
supernovae that were observed with VLBI ($<2.0\times10^4$\,km\,s$^{-1}$), and would imply 
(if we assume a non-decelerated expansion at least until the epoch of our observations) a 
fractional width of the shocked 
circumstellar region of $\sim$0.5. To obtain this estimate, we assume that the optical-line 
emission comes from a region close to the inner edge of the shocked ejecta (see Chevalier \& 
Fransson \cite{Chevalier1994}). This fractional width is much larger than that found for 
SN\,1993J and those predicted from different models of type II supernovae (Chevalier 
\cite{Chevalier1982}). Considering a decelerated supernova expansion would result in 
even larger fractional-shell width estimates. 

\subsection*{General Remarks}

In short, every model used in our analysis results in a supernova size
much larger than expected from the optical-line velocities of this supernova 
and the expansion velocities found for all supernovae that could be 
imaged with VLBI. The lowest average expansion velocity compatible with our VLBI data is a factor 
of 2 larger than the velocity inferred from optical-line emission. However, despite the apparent 
significance of our measurements (4\,$\sigma$ for the 
flux density and 5\,$\sigma$ for the size), we have obtained only a marginal 
detection of SN\,2008ax with our VLBI observations.

\begin{acknowledgements}
The National Radio Astronomy Observatory is a facility of the National Science Foundation, 
operated under cooperative agreement by Associated Universities, Inc. 
This work has been partially funded by grants AYA2006-14986-CO2-01 and AYA2006-14986-C02-02
of the Spanish DGICYT. K. W. W. thanks the Office of Naval Research for the 6.1 funding 
supporting this research. I. M. V. is a fellow of the Alexander von Humboldt-Stiftung.

\end{acknowledgements}

\end{document}